\documentclass[pre,twocolumn,amsmath,amssymb]{revtex4}

\usepackage{graphicx}
\usepackage{amssymb}
\usepackage{amsmath}
\usepackage{enumerate}

\begin{document}

\title{Homeostatic competition drives tumor growth and metastasis nucleation}

\author{Markus Basan}
\affiliation{Laboratoire Physico-Chimie Curie (CNRS-UMR 168, Universit\'e Pierre et Marie Curie Paris $\rm V\!I$),
Institut Curie Centre de Recherche, 26 rue d'Ulm, F-75248 Paris Cedex 05, France}

\author{Thomas Risler}
\email{thomas.risler@curie.fr}
\affiliation{Laboratoire Physico-Chimie Curie (CNRS-UMR 168, Universit\'e Pierre et Marie Curie Paris $\rm V\!I$),
Institut Curie Centre de Recherche, 26 rue d'Ulm, F-75248 Paris Cedex 05, France}

\author{Jean-Fran\c{c}ois Joanny}
\affiliation{Laboratoire Physico-Chimie Curie (CNRS-UMR 168, Universit\'e Pierre et Marie Curie Paris $\rm V\!I$),
Institut Curie Centre de Recherche, 26 rue d'Ulm, F-75248 Paris Cedex 05, France}

\author{Xavier Sastre-Garau}
\affiliation{D\'epartement de Biologie des Tumeurs, Institut Curie, 26 rue d'Ulm, F-75248 Paris Cedex 05, France.}

\author{Jacques Prost}
\affiliation{Laboratoire Physico-Chimie Curie (CNRS-UMR 168, Universit\'e Pierre et Marie Curie Paris $\rm V\!I$),
Institut Curie Centre de Recherche, 26 rue d'Ulm, F-75248 Paris Cedex 05, France}
\affiliation{E.S.P.C.I., 10 rue Vauquelin, F-75231 Paris Cedex 05, France}

\begin{abstract}
We propose a mechanism for tumor growth emphasizing the role of homeostatic
regulation and tissue stability. We show that
competition between surface and bulk effects leads to the existence
of a critical size that must be overcome by metastases to reach macroscopic sizes.
This property can qualitatively explain the
observed size distributions of metastases, while size-independent growth rates cannot account for clinical and
experimental data. In addition, it potentially explains the observed
preferential growth of metastases on tissue surfaces and membranes
such as the pleural and peritoneal layers, suggests a mechanism underlying the seed and soil hypothesis
introduced by Stephen Paget in 1889 and yields realistic values for metastatic inefficiency.
We propose a number of key experiments to test these concepts.
The homeostatic pressure as introduced in this work could constitute a quantitative,
experimentally accessible measure for the metastatic potential of early malignant growths.
\end{abstract}

\maketitle

\section*{Introduction}
The progression of cancer is a multi-step process. Over 80\% of
malignant tumors are carcinomas that originate in epithelial tissues from where they
invade through the basal membrane into the connective tissue. At some point, subpopulations of cells may detach from the
primary tumor and spread via the bloodstream and the lymphatic system. Some of them give rise to metastases in distant organs.
Metastases account for the majority of patients' deaths due to cancer, and thus understanding the metastatic process is of critical importance.
The metastatic cascade is a very inefficient process, as only one in about a thousand
cells that leave the primary tumor goes on to form a macroscopic secondary tumor.
This property is referred to as  ``metastatic inefficiency'' (Chambers {\it et al.}, 2002; Sahai, 2007). Recent experimental results have shown however
that cell extravasation is highly efficient, namely that over 80\% of the metastatic cells
that are present in the bloodstream manage to enter a distant organ (Luzzi {\it et al.}, 1998; Cameron {\it et al.}, 2000; Zijlstra {\it et al.}, 2002).
Thus, the main contribution to metastatic inefficiency arises from the failure of cancerous cells to grow inside invaded organs.
Metastatic tumors also show preferential growth in different organs
with a distribution that cannot be explained by blood flow patterns alone. Hence, the efficiency of the
metastatic process depends on specific interactions between the
invading cancer cells and the local organ tissues (Fidler {\it et al.}, 2003). This concept, referred to as
the ``seed and soil hypothesis'', was introduced by Stephen Paget as
early as 1889 (Weinberg, 2007): ``the seed''---the metastatic cell---needs to be compatible with ``the soil''---the host tissue---for successful growth to occur 
(Fidler, 2003; Couzin, 2003). Despite this early observation, 
the nature of the interactions controlling both the efficiency of the metastatic process and its
tissue specificity remains a poorly understood aspect of cancer progression even today.

\begin{figure}[h]
\scalebox{0.4}{
\includegraphics{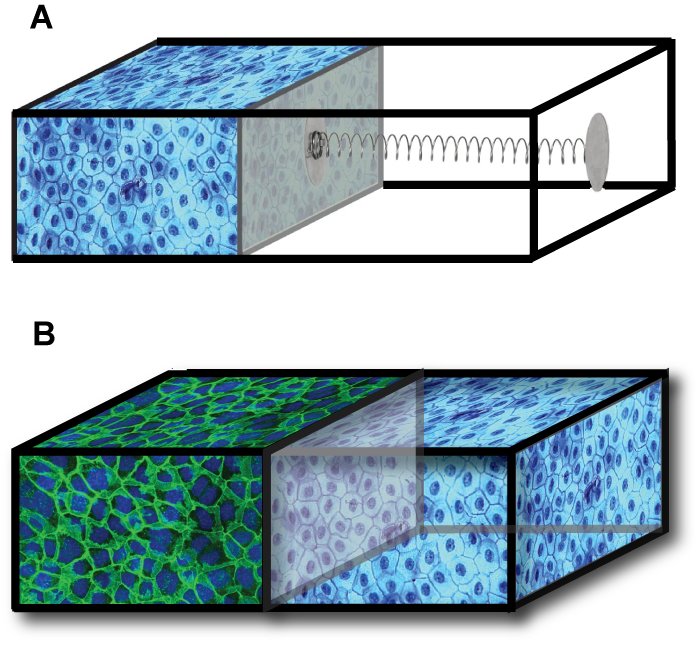}
}
\caption {\label{fig_pressure} (A) Schematic representation of a
measurement apparatus for the homeostatic pressure. As the tissue
proliferates, the piston compresses the spring and the pressure
exerted on the tissue increases. Once a steady state is reached,
cell division and apoptosis balance. The cell density and the
pressure exerted on the spring at this point define the homeostatic mechanical
state of the tissue in a given biochemical environment. (B)
Schematic representation of a tissue-competition experiment. The two
tissues are in mechanical contact through a freely-moving,
impermeable piston. The tissue with the lower homeostatic pressure
is compressed to a cell density above its homeostatic point and
initiates apoptosis. The other tissue proliferates and expands until
the opposing tissue has disappeared.}
\end{figure}

In this work, we introduce the notion of homeostatic pressure and propose that it is
an important property for describing the competition between different tissues grown in a finite volume.
The concept of homeostatic pressure is best defined from the following
experiment: consider a chamber in which cells can be cultured with
a setup that enables successful proliferation, allowing in
particular for water, oxygen, nutrients and growth factors to
diffuse through the compartment walls, keeping the cells'
biochemical environment constant. The compartment is closed
on one side by a piston connected to a rigid wall with a spring
(Fig.~\ref{fig_pressure}A).
 As the growing tissue fills the
available space and gradually compresses the spring, the pressure
rises until a steady state is reached in which division balances
apoptosis and the piston stops moving. The spring position is
stable, since further growth increases the pressure above this value
and favors apoptosis, whereas recession of the piston decreases the
pressure and favors division. This steady state is characterized by
a well-defined pressure exerted on the spring and a well-defined density of cells, 
which we refer to as the homeostatic pressure and density of the tissue in this particular
biochemical environment. Note that the homeostatic pressure is
different from the hydrostatic pressure since the lateral walls of
the chamber allow for fluid transport. Instead, it resembles more an
osmotic pressure but originates from the forces driving tissue
expansion.

We now show that the ability of one tissue to replace another one in a competition for space 
depends on the relative values of their homeostatic pressures.
Let us consider a similar chamber, but in which the
piston separates two tissue compartments $H$ and $T$, establishing mechanical
contact (see Fig.~\ref{fig_pressure}B). Suppose that tissue $H$ has
a homeostatic pressure $p_{H,h}$ smaller than that of tissue $T$,
$p_{T,h}$. As cells divide, the pressure rises and first reaches
the homeostatic pressure $p_{H,h}$. At this point, tissue $H$ stops
growing while tissue $T$ continues to proliferate and drives the
pressure above $p_{H,h}$. As a result, the apoptosis rate of $H$
becomes larger than its division rate, resulting in its recession.
The process continues until tissue 
$H$ completely disappears. The winning
compartment always corresponds to the tissue with the larger
homeostatic pressure.

It is interesting to consider the effect of biochemical
signaling or immunological interactions between the two tissues.
In particular, consider the case where $H$ resists the
expansion of $T$ by locally decreasing the homeostatic pressure of
$T$. If this decrease is large enough, tissue $T$ shrinks and the
result of the competition will be reversed as compared to the case
without signaling. However, if we consider a sufficiently large
compartment $T$, the region close to $H$ has a negligible
contribution to the overall compartment growth, and $T$ expands as
in the absence of signaling. There is a particular size
of compartment $T$ for which, at the homeostatic pressure $p_{H,h}$,
its average growth vanishes: the excess division away from the
piston exactly balances the excess death close to it. A
steady state is possible for this particular size of compartment
$T$, but it is unstable. This introduces the second important
concept of this paper: the existence of a critical size beyond
which a tumor tends to grow and below which it tends to shrink. 

A critical size can also exist due to interfacial tension in
higher-dimensional geometries, such as the two-dimensional
organisation of a monolayered epithelium 
or the three-dimensional configuration of a secondary tumor
within the bulk of a host tissue.
The concept of tissue interfacial tension has already been used to
explain cell sorting of tissues with different adhesive properties
(Duguay {\it et al.}, 2003), and quantified for several tissues (Foty, 1996).
Tissue interfacial tension can also originate from 
the mechanical contraction of cytoskeletal elements at the interface
(Lecuit and Lenne, 2007; Sch\"otz {\it et al.}, 2008).
In a spheroid of tissue $T$ located within the
bulk of tissue $H$, the excess pressure in $T$
is given by Laplace's law: $\Delta p = 2 \gamma/r$,
where $\gamma$ is the interfacial tension between $H$ and $T$ and
$r$ is the radius of the spheroid. As a result, for small
enough radii, the pressure in $T$ is larger than $p_{T,h}$, and
$T$ recedes. For large radii however, the excess pressure as given by Laplace's law
vanishes and we recover the previous one-dimensional case where $T$
grows. There is again an unstable critical radius $r_c$ for which a
steady state exists.

So far, we have considered cell growth and death processes as
entirely deterministic, in which case only tumors larger than the
critical size can grow. However, single cells give rise to tumors
and metastases (Talmadge and Fidler, 1982; Talmadge and Zbar, 1987; Chambers and Wilson, 1988). This is possible because cell growth and death
are stochastic processes. In this paper, we calculate the probability for a single cell to give rise to a macroscopic tumor and
obtain results that are compatible with
experimental data on metastatic inefficiency (Luzzi {\it et al.}, 1998; Cameron {\it et al.}, 2000; Zijlstra  {\it et al.}, 2002). The concepts we use here are
similar to those used to describe the statistics of nucleation
processes as they occur in first-order phase
transitions. It is well known that nucleation is easier on surfaces
or foreign bodies than in the bulk of a system.
The same holds true for tumor growth: we show that it is more
likely for tumors to reach the critical size at an interface than in the bulk of a
tissue, in agreement with experimental and clinical observations (Cameron {\it et al.}, 2000; Weiss, 1985).
Hence, in this paper, we argue that an unstable critical size for tumor growth exists, which is responsible for the inefficiency of the metastatic cascade and
could account for the preferred growth of metastases on surfaces and interfaces. We treat only the early stages of tumor and metastatic growth,
where the heterogeneity of tumors---due to effects such as the diffusion of nutrients and growth factors or genetic mutations---can be neglected.
These effects play an important role for larger tumor sizes only (Hanahan and Weinberg, 2000).

\section*{Results}
\subsection*{Tissue Rheology and Homeostatsis}
While the notion of homeostatic pressure and density is model independent, the details of the tissue dynamics are not.
Here, we employ a continuous description that 
we expect to be valid for systems large compared
to the cell size and time scales large compared to the
characteristic times of individual cellular processes. The local
density of cells $\rho$ obeys the continuity equation:
\begin{equation}\label{Continuity}
\frac{\partial}{\partial t} \rho + \nabla\cdot(\rho \mathbf{v}) = (k_d-k_a)
\rho,
\end{equation}
where $\mathbf{v}$ denotes the local velocity of the tissue and $\nabla\cdot(\rho \mathbf{v})$
the divergence of the cell flux. The right hand side corresponds to source and sink terms that describe
the local production and destruction of cells due to cell division
($k_d$) and apoptosis ($k_a$). In addition, tissues must also
satisfy momentum conservation, which, for systems where inertia plays
a negligible role, reduces to force-balance:
\begin{equation}\label{ForceBalance}
\partial_{\alpha} \sigma_{\alpha\beta} = 0.
\end{equation}
Here, $\partial_{\alpha}$ denotes the partial derivative with respect to the coordinate
$\alpha$ ($\alpha=x,y,z$), and summation over repeated indices is implicit;
$\sigma_{\alpha\beta}$ denotes the total stress tensor that 
we split into a velocity-independent part and a
dynamic part $\sigma'_{\alpha\beta}$. For an isotropic tissue,
the velocity-independent part reads $-p\delta_{\alpha\beta}$, where
$p$ is the tissue pressure discussed above. The viscous part however encodes
the rheological properties of the tissue in a constitutive equation that relates it to
the velocity-gradient tensor $\partial_{\alpha}v_{\beta}$. Tissues are complex
media with a rheological behavior intermediate between those of liquids and
solids (Foty {\it et al.}, 1994; Sch\"otz {\it et al.}, 2008). 
On timescales short compared to their viscoelastic relaxation time, tissues have a finite shear modulus $E$
of the order of $10^2$--$10^4$ Pa (Forgacs {\it et al.}, 1998; Engler {\it et al.}, 2004; Kong {\it et al.}, 2005). For time scales exceeding the largest
relaxation time $\tau$ however, viscoelastic media behave as viscous
liquids with viscosity $\eta=E\tau$. Measurements of the mechanical response of various cell aggregates suggest a value of the relaxation time in the range of tens of seconds to several minutes 
(Forgacs {\it et al.}, 1998; Sch\"otz {\it et al.}, 2008),
corresponding to a viscosity in the range of $10^3$-$10^5$ Pa$\cdot$s. The fastest division rates of mammalian cells are typically of the
order of one division per day (Weinberg, 2007). Hence, tissue-growth dynamics takes place on time scales that are long compared to the characteristic times of cellular processes,
including adhesion and detachment of the proteins that insure the integrity of the tissue under consideration. Under such conditions, it is a general result that the effective rheology on large scales appears to be that of a fluid (Frisch {\it et al.}, 1985).
We therefore argue that, in the context of tissue-growth dynamics, a purely viscous rheology is appropriate, which leads to
the standard constitutive equation:
\begin{equation}\label{Constitutive}
\sigma'_{\alpha\beta} = \eta\,\left( \partial_{\alpha}v_{\beta}+\partial_{\beta}v_{\alpha} \right).
\end{equation}

Under fixed biochemical and biophysical conditions, division and apoptosis rates---as well as pressure---are functions of cell density only. 
In the absence of a detailed knowledge of the
pressure and rate dependences as functions of $\rho$, and for the
sake of simplicity, we rely on an expansion to first order in
$\rho-\rho_h$ around the homeostatic density $\rho_h$:
\begin{eqnarray}\label{EqStateRates}
p&=&\chi^{-1} (\rho-\rho_h)+p_h\nonumber\\
k_d-k_a&=&-\kappa (\rho-\rho_h).
\end{eqnarray}
The parameter $\chi$ is equivalent to the standard compressibility of a material
and describes the variation of cell density with pressure.
Similarly, the coefficient $\kappa$ quantifies how the difference between division and apoptosis rates depend on density.
Both $\chi$ and $\kappa$ are experimentally accessible parameters that must both be positive to insure stability.
In Eqs.~(\ref{EqStateRates}), the expansion of the pressure $p$ in terms of the cell density $\rho$ is complementary to the expansion of $k_d-k_a$ known 
as logistic growth, which is a common way to model growth dynamics (Sachs {\it et al.}, 2001). As we have stated, the most general dependence of the pressure $p$ as well as 
of the division and apoptosis function $k_d-k_a$ on the biochemical environment of the tissue is encoded in the expansion coefficients $\chi$ and $\kappa$ here, which are therefore 
constants only under fixed biochemical conditions. As we focus here on tumors of small sizes under steady environmental conditions, such a dependence will not be discussed here.
But our framework in principle allows to study more complex situations where spatio-temporal inhomogeneities would play a role, simply by allowing $\chi$ and $\kappa$ to vary.
Cell division and apoptosis could also be coupled directly to pressure, leading to tissue competition as proposed in a similar model by B. Shraiman (Shraiman, 2005).
Also, studies have shown that a defective density sensing of cancerous cells can lead to a growth advantage (Chaplain {\it et al.}, 2006). 
In this paper, we argue that competition for volume is a generic property of tissues in mechanical contact, since the pressure and effective growth rate functions
take the form of Eqs.~(\ref{EqStateRates}) close to the steady state density of the tissue.

\subsection*{Tumor Growth Dynamics}
\begin{figure}[h]
\scalebox{0.33}{
\includegraphics{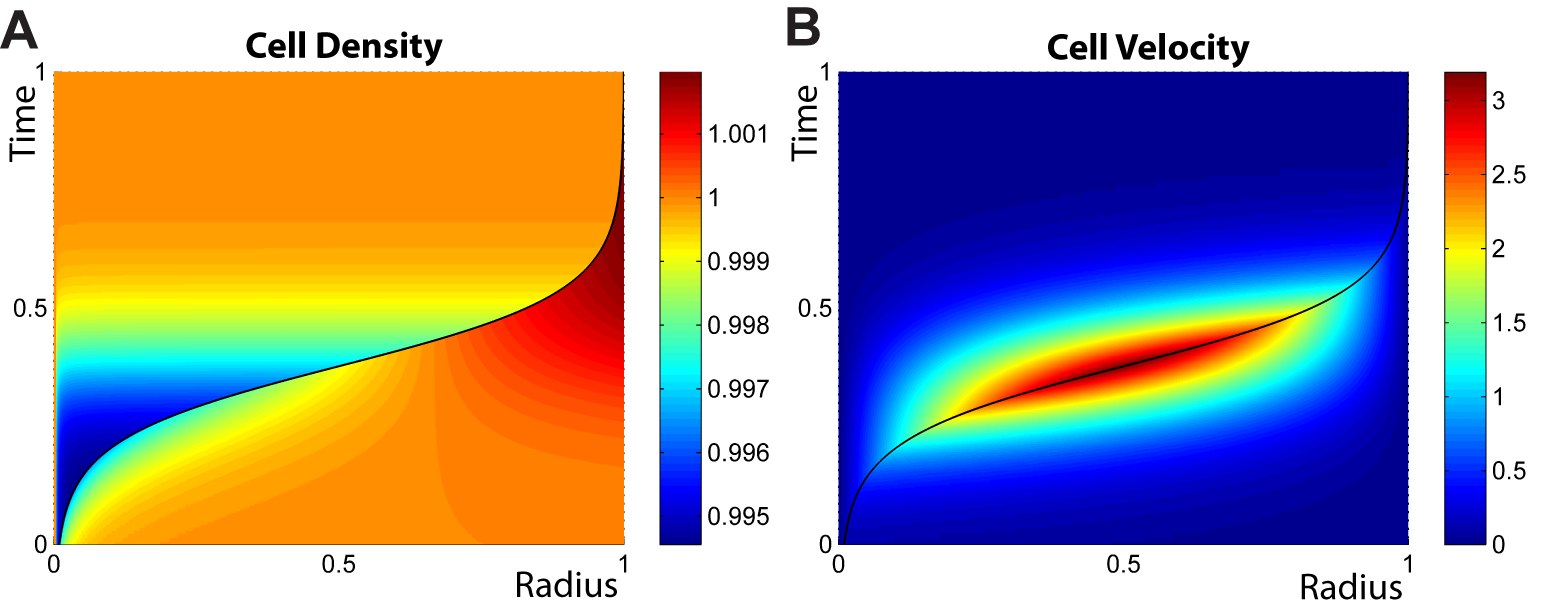}
}
\caption {\label{fig_spherical} Numerical solution for the cell
density (A) and cell velocity (B) as functions of space and time
during the growth of one tissue located in the bulk of another
tissue of lower homeostatic pressure.
Color coding for local cell density and velocity is given on the right hand side of panels (A) and (B), respectively.
Spherical symmetry is assumed.
Total integration time, compartment size and homeostatic densities of tissues $T$ and $H$
are scaled to one. In both plots, the boundary between the two
tissues is indicated by a black line. Parameters are chosen in order
to illustrate the interplay between viscous dynamics and compartment
growth (see supplementary material).}
\end{figure}

We are now in a position to show that homeostatic regulation
intrinsically contains a growth mechanism for neoplastic tissues.
Consider the growth of a spherical tissue $T$ located at the center
of a spherical compartment of finite volume, filled with another
tissue $H$ of lower homeostatic pressure.
The spherical tissue can be either a primary tumor developing within the tissue it stems from, or a metastasis that has migrated from 
its original location and invaded a foreign organ. The two tissues are in mechanical contact, so that the total stress is continuous at the interface.
Eqs.~(\ref{Continuity})-(\ref{EqStateRates}) must be solved for both compartments, taking the location of the interface of the two tissues into account.
A numerical solution of the associated generic growth dynamics is presented in Fig.~\ref{fig_spherical}.
The solution shows that the tissue with higher homeostatic pressure
grows at the expense of the other one and takes over the entire compartment.
{\it In vivo} however, the condition of a fixed finite volume does not hold in general. In real tissues, there is often first a displacement of the non tumor tissue 
before anatomical constraints limit the total volume available to the system. However, the devastating effect of malignant tumors stems from the fact that they invade and replace the functional tissues. The architecture of most tissues leads to a competition for volume in the case of neoplastic proliferation. Note that, considering the time evolution of the boundary between the two tissues only, we get a curve that is reminiscent of the well-kown, experimentally-observed Gompertzian growth curves (Molski and Konarski, 2003). A quantitative illustration of this behavior obtained within our framework is illustrated in the supplementary material, making use of realistic parameters.

\begin{figure}[h]
\scalebox{0.33}{
\includegraphics{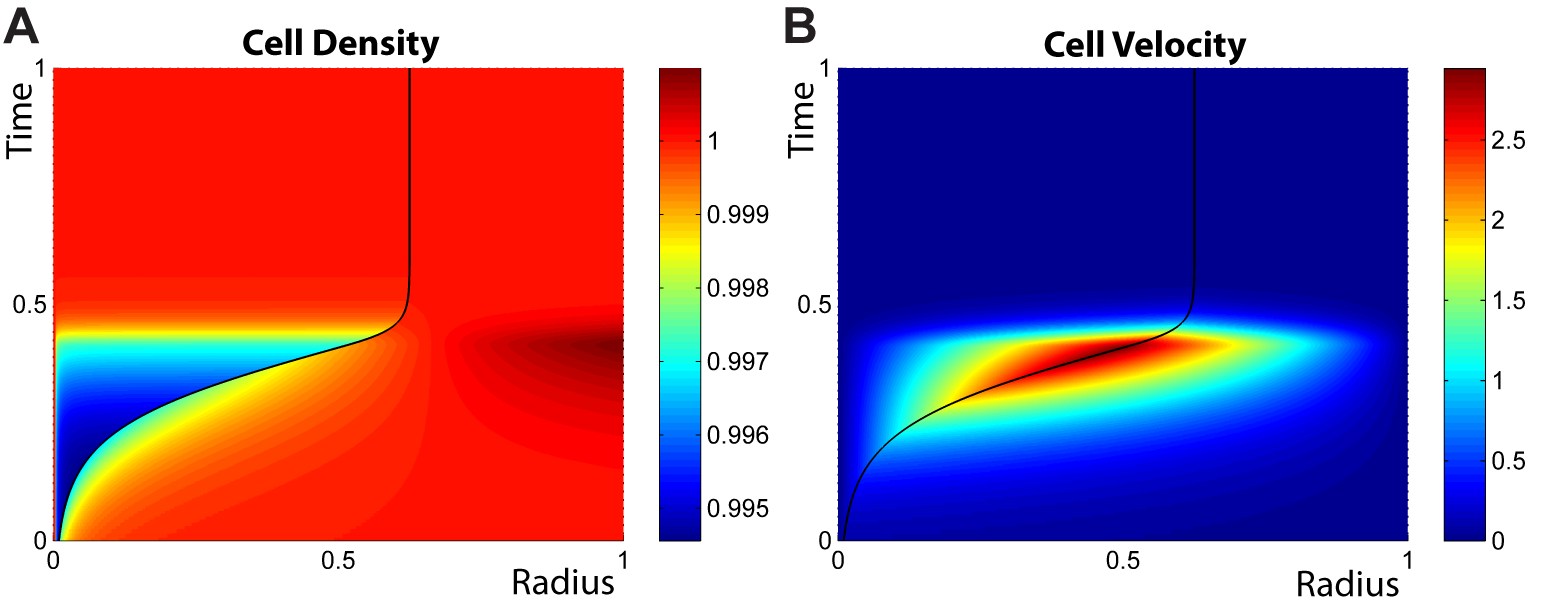}
}
\caption {\label{fig_membrane} Numerical solution for the cell
density (A) and cell velocity (B) as functions of space and time
during the growth of a tissue engulfed in an elastic membrane, and
located in the bulk of another tissue of lower homeostatic
pressure. As in Fig.~\ref{fig_spherical}, spherical symmetry is assumed; total integration time,
compartment size and homeostatic densities are scaled to one, and in
both plots the boundary between the two tissues is indicated by a
black line. Color coding is similar to the one used in Fig.~\ref{fig_spherical}.
In this solution, the membrane is treated as purely elastic and is put under
tension above a given radius $x_{0}=0.5$. The explicit surface tension dependence on the boundary location $x$
is given in the supplementary material. Additional parameters are given in the supplementary material.
(A) The expansion of the inner compartment is similar to that of Fig.~\ref{fig_spherical} until, at $x=0.5$,
surface tension begins to play a role. The membrane expands until its tension balances the pressure difference
between the two compartments and a stable steady state is reached.
(B) Corresponding cell-velocity plot.}
\end{figure}

We now examine several effects that can significantly alter
the tumor growth dynamics as presented above.
A first example, which is motivated by the structure of benign tumors,
corresponds to tissue $T$ engulfed in a membrane, typically a thin shell of extracellular matrix, where the surface
tension $\gamma$ rises as $T$ expands. If this tension
increases faster than the radius of $T$, the additional
pressure increases and the expansion of $T$ eventually stops: a
stable steady state exists at $r=r_s$ such that
$p_{T,h}-p_{H,h}=2\gamma(r_s)/r_s$, where $p_{T,h}$ and $p_{H,h}$ are the
respective homeostatic pressures of $T$ and $H$.
A numerical solution illustrating this
case is presented in Fig.~\ref{fig_membrane}.
This dormant state is stable until genetic alterations inducing the production of proteases by the tumor cells
lead to the degradation of the membrane.

\begin{figure}[h]
\scalebox{0.33}{
\includegraphics{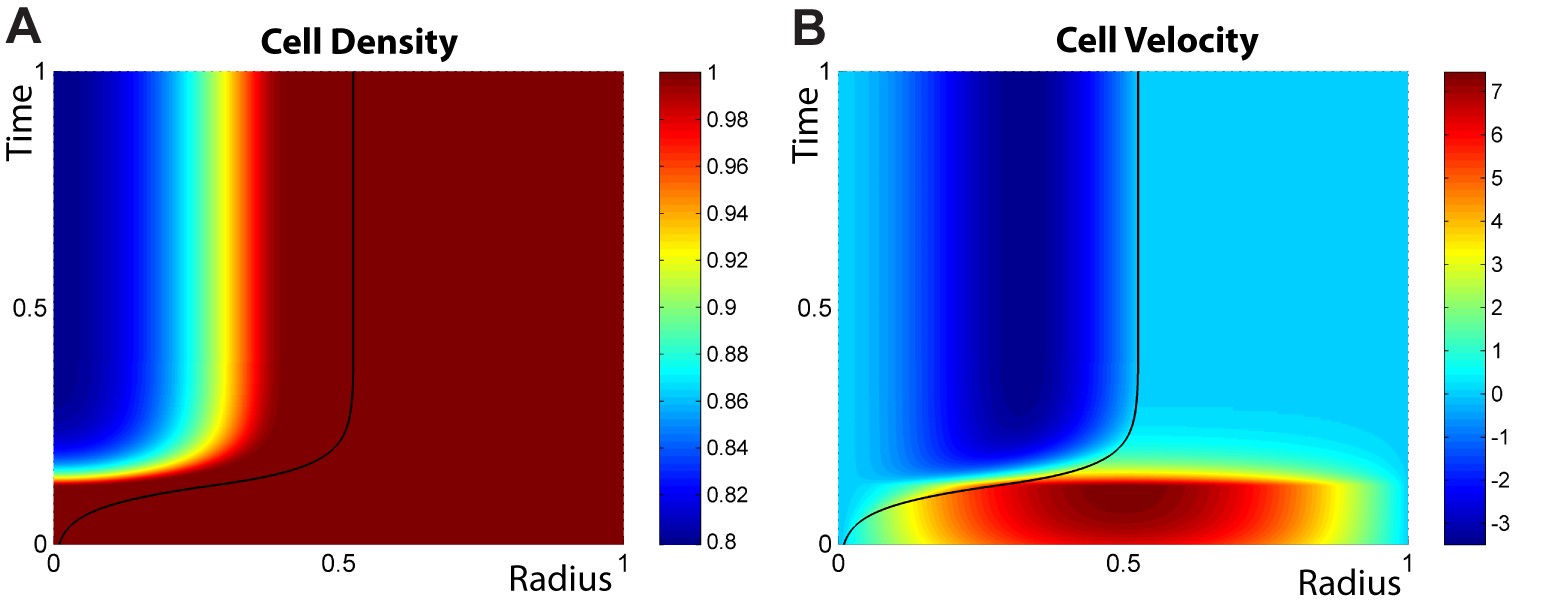}
}
\caption {\label{fig_nutrient} Numerical solution of the growth
dynamics with the geometrical arrangement of Fig.
\ref{fig_spherical} when growth rates are nutrient-limited.
Nutrients diffuse into the inner compartment through the tissue
interface (see supplementary material). (A) The inner compartment
starts growing as in the case of Fig.~\ref{fig_spherical} but
asymptotically reaches a maximum size.
(B) Cells proliferate at the surface of the 
inner compartment which is rich in nutrients and die at the
center where nutrients are scarce, resulting in an inward flow of
cells. Color coding is similar to the one used in Fig.~\ref{fig_spherical} 
and~\ref{fig_membrane}, but it  now allows for negative values required by the inward flow.}
\end{figure}

As a second example, we consider the case of a tumor that is limited in its growth, for example by nutrient or oxygen supply.
It is indeed a well-known fact that tumors are poorly vascularized before they acquire the capability to
trigger the growth of new blood vessels via angiogenesis (Folkman and D'Amore, 1996; Weidner, 1991; Hanahan and Weinberg, 2000).
This limitation has profound consequences for their growth dynamics (Preziosi, 2003), often leading
to the existence of a maximum size of about one to two millimeters,
where they remain in a ``dormant state'' until the induction of
angiogenesis (Folkman and D'Amore, 1996; Weinberg, 2007). In
Fig.~\ref{fig_nutrient}, we present a numerical integration of the
growth dynamics of a nutrient-limited
tumor in the bulk of a healthy, well vascularized tissue.
 While we
assume a homogeneous and high 
enough concentration of nutrients for the
healthy tissue, the neoplastic tissue is only supplied with nutrients via
diffusion through its surface. Since nutrient diffusion is fast
compared to growth dynamics, we calculate the nutrient concentration
profile by solving a steady-state diffusion equation, taking into
account nutrient consumption due to cell metabolism and division
(see supplementary material). We choose functional dependences of
the division and apoptosis rates on the nutrient concentration and
cell density that correspond to a biological behavior: under very 
low concentrations of nutrients or oxygen, cells tend to die,
but a limited supply of nutrients can also decrease cell division
by triggering cell differentiation, inducing a quiescent cell state or favoring adaptation of the metabolism of the cells
to the new environment. In agreement with what is known 
about the internal structure of dormant tumors, cells divide at the
boundary where they get enough nutrients, and die at the center.
This creates a steady state flow of cells from the surface toward
the center of the tumor and thereby a constant cell turnover that
is favorable to mutations.

Note that the typical length-scale at which the diffusion of
nutrients becomes a limiting factor is of the order of millimeters,
the size of a dormant tumor (Folkman and D'Amore, 1996; Weinberg, 2007).
This scale is very large compared to the size we estimate for the
critical radius introduced above. In the following, when considering
the nucleation process of micro-tumors, we therefore assume a
homogeneous, high enough concentration of nutrients.

\subsection*{Critical Size and Stochastic Growth Dynamics}
\begin{figure}[h]
\scalebox{1.5}{
\includegraphics{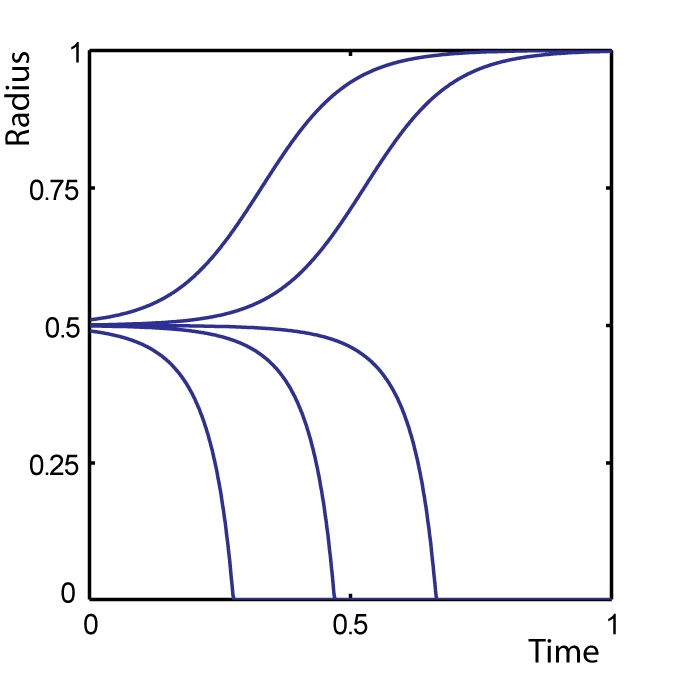}
}
\caption {\label{fig_critical} Tissue boundary as a function of time
during the growth of a tissue located in the bulk of another tissue of
lower homeostatic pressure, with interfacial tension and in
spherical geometry. Parameters are given in the supplementary material,
together with $r_c=0.5$ and $\gamma =1$. The different curves show the dynamics for the following initial
values $r_0$ of $r$: 0.49, 0.499, 0.4999, 0.501 and 0.51. Both
tissues start out at their homeostatic densities.}
\end{figure}

A third effect that can modify tumor growth dynamics is the presence
of a constant tissue interfacial tension as introduced above.
When $H$ and $T$ are at their homeostatic densities, there exists a
particular radius $r_c$ at which mechanical
equilibrium is reached, but this equilibrium, given by $p_{T,h}-p_{H,h}=2\gamma/r_c$, is unstable.
Numerical solutions illustrating the growth dynamics around this critical radius are
presented in Fig.~\ref{fig_critical}.

Given the existence of such an unstable critical radius,
the question arises as to how a metastasis---or a primary tumor---can grow
within a healthy tissue since in general it originates
from a single cell (Talmadge and Fidler, 1982; Talmadge and Zbar, 1987; Chambers and Wilson, 1988). The answer stems from stochasticity, an aspect of
the dynamics that has been ignored in the description so far.
The importance of stochasticity in growth processes has already been recognized in various situations (Nowak {\it et al.}, 2003).
Under the assumption that stochastic tumor growth is a
Poisson process, the evolution of the probability $P(n,t)$ for a spherical
tumor $T$ inside a healthy tissue $H$ to contain $n$ cells at time
$t$ can be described by a master equation:
\begin{eqnarray}\label{MasterEquation}
\frac{d P(n,t)}{dt} &=& r^{+}_{n-1} P(n-1,t) - (r^{+}_{n} + r^{-}_{n}) P(n,t) \nonumber\\
&& + r^{-}_{n+1} P(n+1,t),
\end{eqnarray}
where $r^+_{n}=nk_d$ and $r^-_{n}=nk_a$ are the rates at which a
tumor grows or shrinks from $n$ to $(n+1)$ or $(n-1)$ cells,
respectively. The rates $k_d$ and $k_a$ depend on $n$,
and we model their dependence in the following way:
For tumors small compared to the size of the
healthy compartment, the healthy tissue is only slightly perturbed away from
its homeostatic state. Thus, the pressure inside the tumor is
given by Laplace's law: $p_{T}=p_{H,h}+2\gamma/r$.
Therefore, the division and apoptosis rates of a spherical tumor of radius $r$ are given by:
\begin{equation}
k_{d/a} = -\kappa_{d/a}\,\chi_T\left( p_{H,h}+\frac{2 \gamma}{r}-p_{T,h}\right) + k_0.
\end{equation}
Here, $\kappa_d$, $\kappa_a$ and $k_0$ are three phenomenological coefficients that enter
the linear expansions of $k_d$ and $k_a$, similarly to $\kappa$ in Eq. (\ref{EqStateRates}).
To ensure the proper behavior as a function of the cell density $\rho$, $\kappa_d$ needs to be positive and $\kappa_a$ negative.
Both equations for $k_d$ and $k_a$ share the same constant $k_0$ such that Eq. (\ref{EqStateRates}) is satisfied with $\kappa= \kappa_d-\kappa_a$.
In the master equation (\ref{MasterEquation}), the rates $r^+_{n}$ and $r^-_{n}$ are then given to leading order by:
\begin{eqnarray}\label{RatesStochastic}
&& r^{+/-}_n = \nonumber\\
&& n\left[ -\kappa_{d/a}\,\chi_T\,\left( p_{H,h}+2
\gamma\left( \frac{4\pi\rho_{T,h}}{3n}\right)^{1/3}-p_{T,h} \right)
+ k_0 \right],\nonumber\\
\end{eqnarray}
and $k_0>0$ fixes the amount of cell turnover---and thereby the amount of stochasticity---in the system.

\begin{figure}[h]
\scalebox{0.2}{
\includegraphics{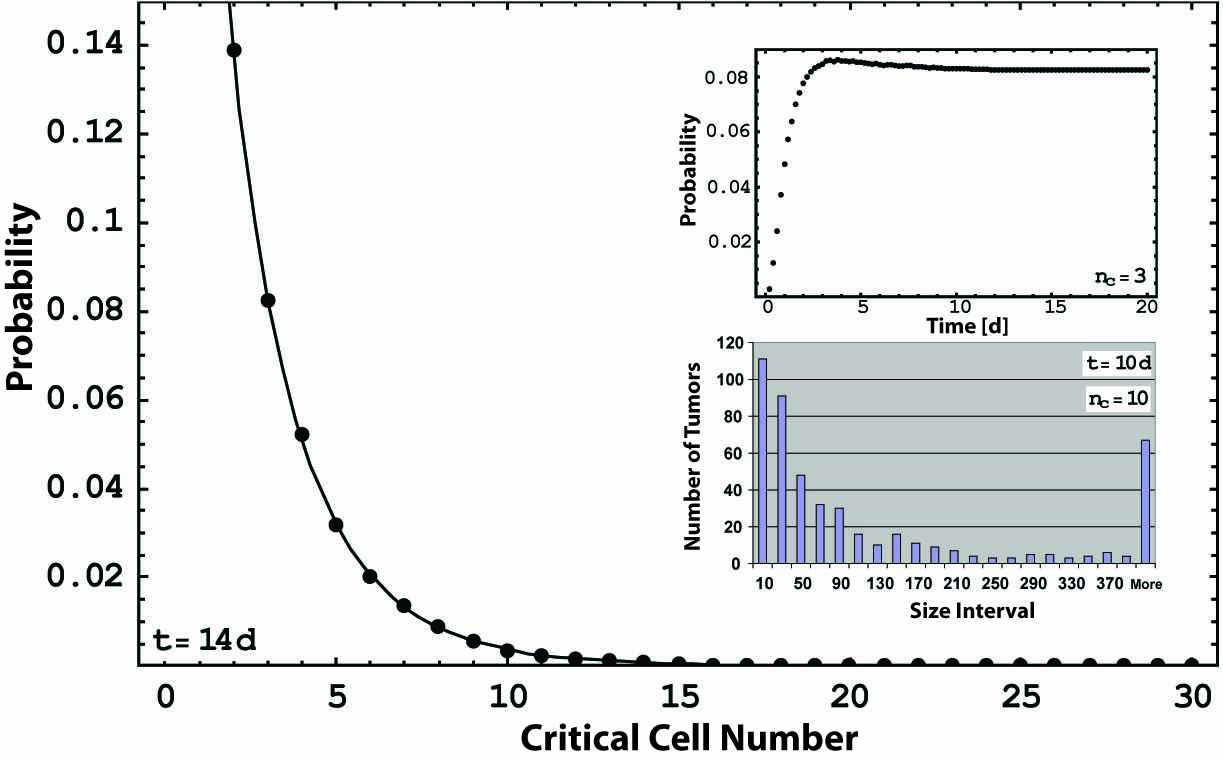}
}
\caption {\label{fig_metastatic} Splitting probability as a function
of the critical cell number. The black dots are the fraction of
tumors with a size larger than the critical radius after 14 days of
evolution, starting from a single cell 
in a Monte Carlo simulation (see Eqs. (\ref{MasterEquation}) and (\ref{RatesStochastic})).
The black line represents the
analytic result for the splitting probability as given by Eq.
(\ref{metastatic_eqn}). Parameters are chosen such that far above
the critical radius the division rate reaches a value of one
division per day and the apoptosis rate is negligible. $k_0$ in Eq. (\ref{RatesStochastic})
is chosen to be 0.9 division per day. Note that the
probability for growth on the compartment surface can be read out
from the plot using half of the critical cell number of the process
in the bulk. Upper inset: Fraction of tumors that are above the
critical size. This fraction relaxes to a constant value in a Monte
Carlo simulation. Lower inset: Distribution of tumor sizes in a
Monte Carlo simulation after 10 days of evolution of $10^5$ individual tumor cells.}
\end{figure}

For an analytic treatment, we map this
growth process onto a random walk with sinks at $n=0$ and $n=n_{\rm{max}}$, which results in
a linear birth-death process where all tumors either disappear or reach
macroscopic sizes when time goes to infinity. The
so-called ``splitting probability'' $\Pi_g$---namely the probability for a single cell to reach the size $n=n_{\rm{max}}$ and not
disappear in the lower sink $n=0$---is given by (Van Kampen, 2007):
\begin{equation} \label{metastatic_eqn}
\Pi_g = \frac{1}{1 + \sum_{\mu = 1}^{n_{\rm{max}} -1} \frac{ r^{-}_{\mu} r^{-}_{\mu-1} .. . r^{-}_{1}}{ r^{+}_{\mu} r^{+}_{\mu-1} .. . r^{+}_{1}} } .
\end{equation}
In Fig.~\ref{fig_metastatic}, we present this analytic result together with the results of a Monte Carlo simulation 
of Eqs.~(\ref{MasterEquation})-(\ref{RatesStochastic}) based on a Gillespie algorithm (Gillespie, 1977).

Finally, it is interesting to consider the growth of neoplastic semi-spheroids on tissue boundaries.
The internal pressure of a semi-spheroid is again given by Laplace's law, but for
the same radius, the number of cells within the tumor is half that
of a complete spheroid. Therefore, the critical number of cells of a
tumor at an interface is only half that of a tumor in the bulk. We
can directly read the resulting effect on the growth probability out
of Fig.~\ref{fig_metastatic}. For example, between the critical
sizes of five and ten cells, we obtain a growth probability ratio of
8.6. This characterizes a significant preference for tumors to grow on
surfaces, an effect that has been observed experimentally
(Cameron {\it et al.}, 2000) and clinically (Weiss, 1985).
Note that other mechanisms---such as the adhesion of cancerous cells to the extracellular matrix, 
which leads to a different initial distribution of metastatic cells---might also play a role.

\section*{Discussion}
In this work, we have shown that homeostatic regulation of cell
density and pressure leads to a competition for space between
tissues in mechanical contact. We have proposed that an increased
homeostatic pressure is a characteristic trait of tumors.
Gedanken experiments for measuring the
homeostatic pressure and growth rates have been discussed
and could be realized using current experimental techniques.
For example, Helminger et al. have shown that tumors
growing in agarose gels proliferate until the pressure exerted by the
gel reaches 45-120 mmHg (Helminger {\it et al.}, 1997). This
gives an estimate of the homeostatic pressure defined in this
paper. Such numbers are compatible with the pressure typically generated
by actin polymerization (Footer {\it et al.}, 2007; Marcy {\it et al.}, 2004).
An experiment of particular interest that has not been performed so far
is the competition between a healthy and a tumor-like tissue separated by a piston,
which could prove that mechanical effects are important for tumor growth.
If the growth mechanism discussed in this paper is relevant to 
tumor growth, it would be interesting to measure the homeostatic pressures characterizing
healthy and neoplastic tissues, together with their dependences on
their biochemical environments. Of particular interest would be 
their dependence on oxygen, nutrients, growth factors and drugs.

The second concept introduced in this paper is the existence of a
critical size for tumor growth due to biochemical, immunological or
mechanical surface effects that can outbalance the bulk growth
advantage of the neoplastic tissue for small tumor sizes. We show
that this interaction can be responsible for the inefficiency of the
metastatic cascade after extravasation. The growth of very few
metastases to macroscopic sizes cannot be explained by
size-independent growth rates, which yield a probability
distribution of metastatic cell clusters that decays exponentially with 
cluster size. Instead, the existence of a critical size
yields realistic values for metastatic inefficiency and a distribution of
tumor sizes compatible with experimental observations
(Luzzi {\it et al.}, 1998; Cameron {\it et al.}, 2000; Zijlstra  {\it et al.}, 2002) (see
Fig.~\ref{fig_metastatic}). Fig.~\ref{fig_metastatic} also shows the dependence of
metastatic inefficiency on the critical size:
a small change in tissue-tumor interaction such as an increased
interfacial tension can dramatically lower the probability for
macroscopic growth. As an illustration of this effect, consider a
metastatic tissue with a critical cell number of 5 in a given
environment. Let us compare this situation with that of the same
tissue placed in another environment where its interfacial tension
is now twice as large, a situation that is well within natural
variations (Foty, 1996). While for the first environment,
about 3 in 100 metastatic cells form a macroscopic tumor, in the
second environment, with a critical cell number of 40, less than 2 
in ten million manage to do so. This corresponds to a difference in
metastatic efficiency of five orders of magnitude. We propose that this
effect could account for the strong tissue specificity of metastatic
growth that underlies the ``seed and soil hypothesis'' (Weinberg, 2007).

The concept of homeostatic pressure presented here is not an alternative
to the cellular and genetic mechanisms involved in tumor growth, but
rather a different level of description. Indeed, we propose that
some of the fundamental biological deregulations that are
characteristic of neoplastic cells lead to an
increased homeostatic pressure. The framework presented here can be used
to explicitly take into account such well-known properties.
It can also be generalized to incorporate more general features of biological tissue behavior.
For example, on long time scales, genetic instability as well as senescence render tissue properties time
dependent. This could be incorporated into our
framework using techniques similar to those of Hallatschek et al.
(Hallatschek {\it et al.}, 2007; Hallatschek and Nelson, 2008),
as well as those of multiscale models of tumor growth (Ribba {\it et al.}, 2006; Macklin and Lowengrub, 2007; Wise {\it et al.}, 2008).

\begin{acknowledgments}
We thank M. Bornens, M. Piel and P. Silberzan for many helpful discussions, and P. Janmey for useful comments.
\end{acknowledgments}

\newpage

\begin{center}
\textbf{\Large{Supplemental Material}}
\end{center}

\section{Spherical Growth}

As an example of a tissue growth competition for which we can solve
the complete dynamics given by the Eqs.~\textbf{1}-\textbf{4} of the main
text, we examine the growth of a spherical tissue located in the
center of a spherical compartment filled with another tissue of
lower homeostatic pressure and enclosed by a rigid boundary. For
now, we neglect all surface tension effects. The force balance
condition (Eq.~\textbf{2}, main text) in three dimensions takes the
form:
\begin{eqnarray}\label{forceBalance}
\frac{\partial \sigma'_{rr}}{\partial r} + \frac{2(\sigma'_{rr}-\sigma'_{\theta
    \theta})}{r} &=& \frac{\partial p}{ \partial r}\nonumber\\
\sigma'_{\phi \phi} &=& \sigma'_{\theta \theta},
\end{eqnarray}
where $r$, $\theta$ and $\phi$ are the spherical coordinates,
while the constitutive equation (Eq.~\textbf{3}, main text) reads:
\begin{eqnarray}\label{constitutive}
\sigma'_{rr} &=& 2 \eta \frac{\partial v}{\partial r}\nonumber\\
\sigma'_{\theta \theta} &=& 2 \eta \frac{v}{r}.
\end{eqnarray}
The continuity equation (Eq.~\textbf{1}, main text), together with
the expansion of $k_d-k_a$ to first order in $\rho-\rho_h$ (Eq.~\textbf{4}, main text), gives:
\begin{equation} \label{cont}
\frac{\partial \rho}{\partial t} + \frac{1}{r^2}
\frac{\partial}{\partial r} \left( r^2 \rho v \right) = - \kappa (\rho-\rho_h)\rho.
\end{equation}
These equations need to be solved for the whole system composed of
the two tissues, together with the moving boundary between them.
Boundary conditions are composed of two parts: (a) in the center and
at the rigid external wall, the velocity field vanishes, such that
$v(r=0) = v(r=R) = 0$; (b) at the interface of the two tissues, the
velocity field and the stress tensor are continuous. The continuity
of the velocity field at the interface leads to the following
equation for the time-dependent location $x(t)$ of the tissue
boundary:
\begin{equation} \label{bound}
\frac{\partial x}{\partial t} = v(x).
\end{equation}

The force-balance condition (\ref{forceBalance}) can be integrated using:
\begin{equation}
\frac{\partial\sigma'_{\theta \theta}}{\partial r} = \frac{1}{r} (\sigma'_{rr}-\sigma'_{\theta \theta})
\end{equation}
from the constitutive equation (\ref{constitutive}) to give:
\begin{equation} \label{simp}
2 \eta \bigg( \frac{\partial v}{\partial r} + \frac{2 v}{r} \bigg) = p - p_{\rm{ext}}.
\end{equation}
In the absence of surface tension, the integration constant
$p_{\rm{ext}}$ is the external pressure imposed by the rigid wall to
satisfy the boundary condition of vanishing velocity. The growth
dynamics consisting of Eqs.~(\ref{cont}) and (\ref{simp}) for each
of the two compartments---together with the moving boundary condition
Eq.~(\ref{bound})---can be solved numerically using a
finite-difference method (Press {\it et al.}, 1992). Results with the
parameters of Table \ref{parametersSpherical} are displayed in
Fig.~\textbf{2} of the main text and show how the inner tissue takes
over the whole compartment.

In the main text, a constant interfacial tension is introduced
between the two tissues. It is shown that this effect leads to the
existence of an unstable critical radius in spherical geometry.
The growth dynamics with an unstable critical radius is
illustrated in Fig.~\textbf{5} of the main text. However, biologically
relevant situations may involve tissues enclosed in membranes whose
tensions increase as the inner tissue grows. This is for example the
case for some benign tumors that undergo growth arrest due to the
extracellular membrane engulfing them. In that case, the surface
tension is now dependent on the location $x$ of the boundary between
the two tissues. For a purely elastic membrane that is put under
tension above a given radius $x_0$, we have:
\begin{equation} \label{membrane}
\gamma(x) = \gamma_0\,\frac{(x-x_0)^2}{x_0^2}\,\theta(x-x_0),
\end{equation}
where $\theta(x)$ is the heaviside step function. A numerical
solution of the growth dynamics with this type of surface tension is
presented in Fig.~\textbf{3} of the main text
(parameters are given in Table \ref{parametersSpherical}
together with $\gamma_0=5$ in scaled units).

\section{Tumor Growth Dynamics with Realistic Parameters}

\begin{figure}[h]
\scalebox{0.19}{
\includegraphics{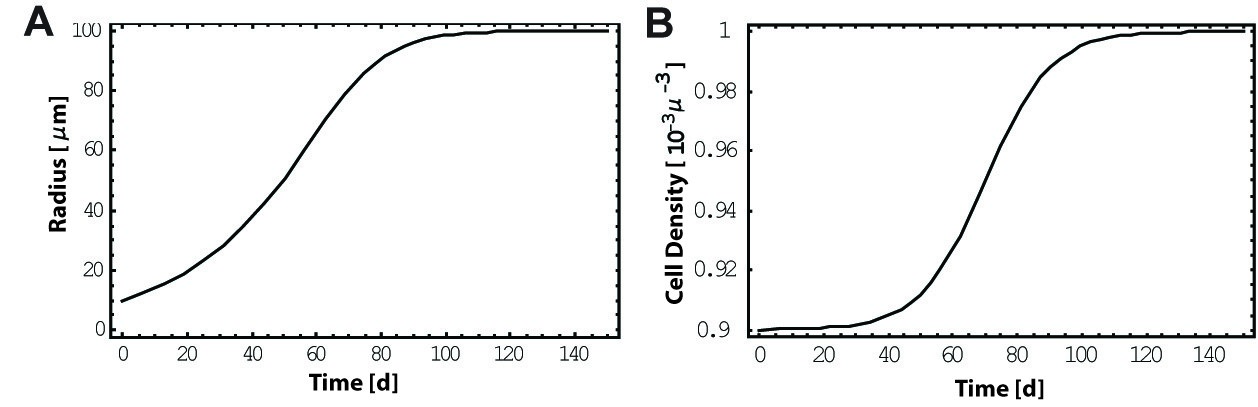}
}
\caption {\label{fig_parameter} Radius and cell-density as functions
of time for a growing tumor in conditions identical to those of
Fig.~\textbf{2} of the main text, but with estimates of realistic
parameters. Cell density is one per $10^3$ $\mu$m$^3$, and cell
division rate is one per day. The homeostatic pressure difference
and the tissue viscosity are $10^3$ Pa and $10^4$~Pa$\cdot$s,
respectively (Helminger {\it et al.}, 1997; Kruse  {\it et al.}, 2000; Forgacs  {\it et al.}, 1998).
Tissue compressibility is $10^{-7}$~Pa$^{-1}\cdot$m$^{-3}$
(Tschumperlin  {\it et al.}, 2004). With these parameters, a separation of
timescales between the slow cell division and the comparatively fast
viscous dynamics occurs, such that we can assume homogeneous cell
densities in both compartments. Starting from a single cell and
neglecting nutrient coupling, a tumor needs about 100 days to fill a
compartment with radius $100$$\mu$m.}
\end{figure}

The growth rate of tumor cells is possibly very slow compared to the
viscous relaxation time. In such a case, it can be assumed that the
cell density in each compartment is constant. In the absence of
surface tension, the pressures in the two compartments balance,
leading to an equation relating the two densities:
\begin{equation}
\chi_H^{-1}(\rho_H-\rho_{H,h}) + p_{H,h}= \chi_T^{-1}(\rho_T - \rho_{T,h}) + p_{T,h}.
\end{equation}
The change in density in each compartment has contributions coming
from the total cell division and apoptosis taking place in the
compartment, as well as from the movement of the boundary $x(t)$:
\begin{eqnarray}
\dot{\rho_T} &=& -\kappa_T\left(\rho_T-\rho_{T,h}\right)\rho_T - 3\,\frac{\dot{x}}{x}\,\rho_T\nonumber\\
\dot{\rho_H} &=& -\kappa_H\left(\rho_H-\rho_{H,h}\right)\rho_H + 3\,\frac{\dot{x}}{x}\,\frac{x^3}{R^3-x^3}\,\rho_H.
\end{eqnarray}
This system of differential equations can be solved numerically.
Results with the parameters given in Table \ref{parametersRealistic}
are displayed in Fig.~\ref{fig_parameter}.

\section{Nutrient-Limited Growth}

\begin{figure}[h]
\scalebox{0.3}{
\includegraphics{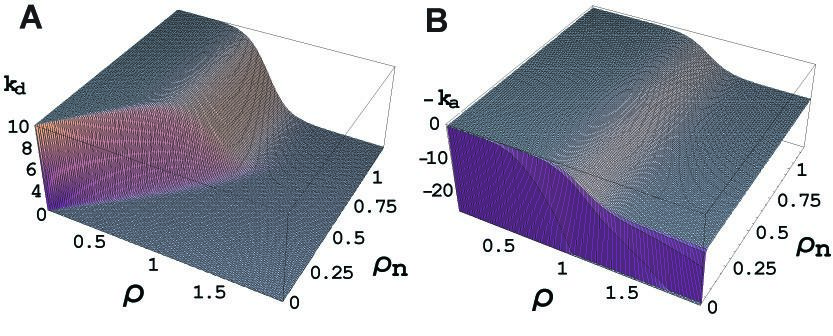}
}
\caption {\label{fig_parametric2} The cell division rate $k_d$ (A)
and the negative apoptosis rate $-k_a$ (B) as functions of the cell
density $\rho$ and the nutrient concentration $\rho_n$, as given by
Eq.~(\ref{rates}). Parameters are given in Table
\ref{parametersNutrients}.}
\end{figure}

\begin{figure}[h]
\scalebox{0.3}{
\includegraphics{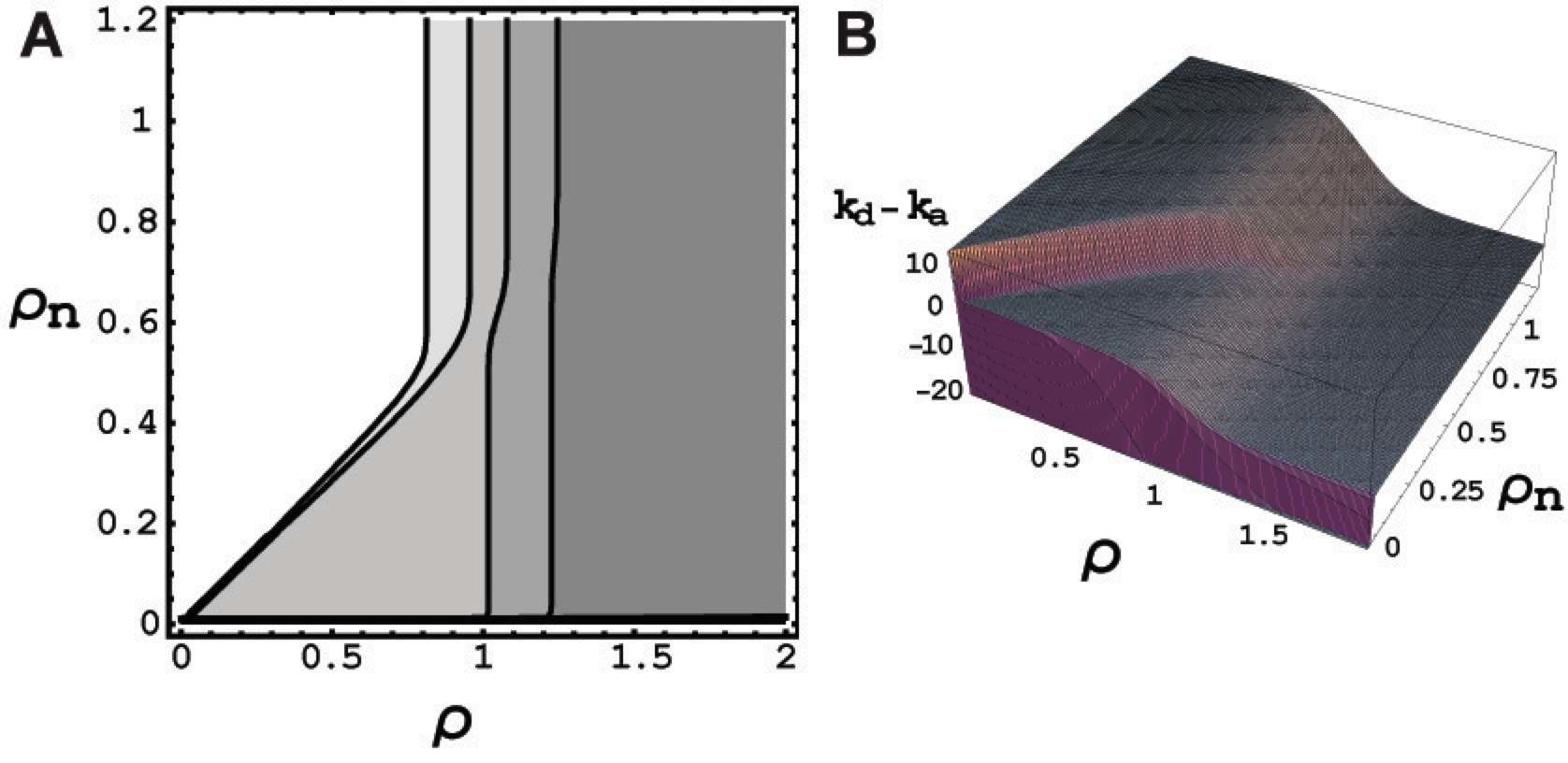}
}
\caption {\label{fig_parametric} Difference $k_d-k_a$ as a function
of cell density and nutrient concentration, as given by Eq.~(\ref{rates}).
Parameters are given in Table \ref{parametersNutrients}. (A) Top
view in grey-scale coding. (B) Three-dimensional view.}
\end{figure}

\begin{figure}[h]
\scalebox{0.3}{
\includegraphics{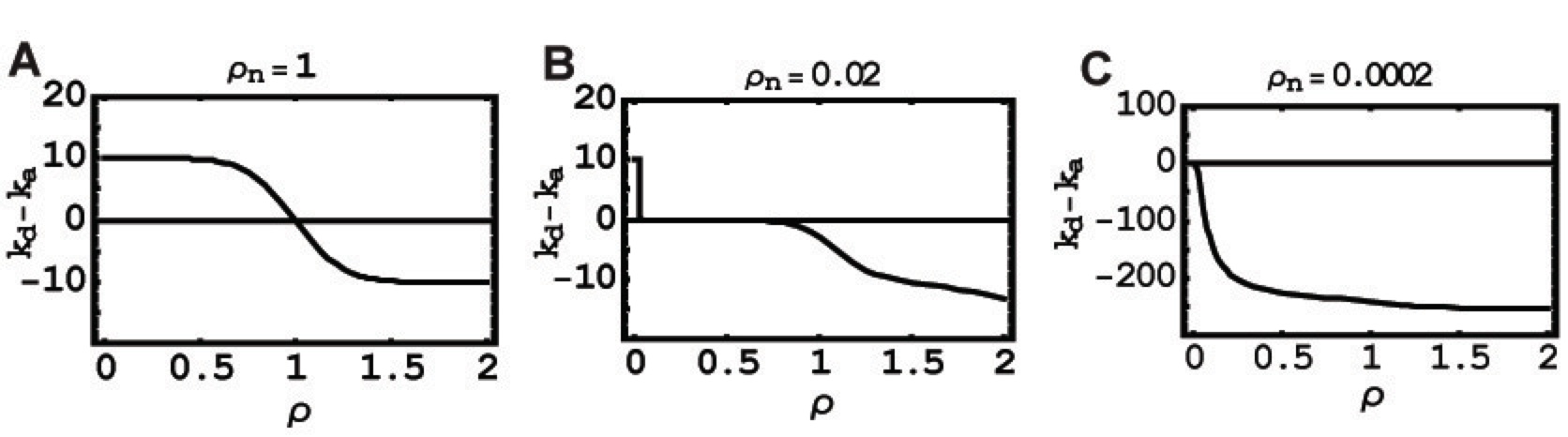}
}
\caption {\label{fig_parametric1} Effect of nutrient concentration
on the difference $k_d-k_a$ as a function of cell density.
Parameters are given in Table \ref{parametersNutrients}. At high
nutrient concentrations, regulation towards the homeostatic density
$\rho_h=1$ is intact, while at low nutrient concentrations cell
division drops significantly. At very low nutrient concentrations,
cells undergo apoptosis at a high rate.}
\end{figure}

When studying the nutrient-limited growth of a tumor prior to
angiogenesis, the dependences of the cell division and apoptosis rates $k_d$ and $k_a$ 
on the nutrient and cell densities $\rho_n$ and $\rho$ are constructed using two assumptions: 
(a) below a given concentration of nutrients per cell $c_1$, cells
stop dividing; (b) below a second, lower concentration of nutrients
per cell $c_2$, cells undergo apoptosis. We model these properties
with the following functions\footnote{Note that the exact form of
these functions is not important. Only their qualitative behavior is
relevant.}:
\begin{eqnarray}\label{rates}
&& k_d = \nonumber\\
&& \frac{\bar{k}_1}{1+\exp\left(\alpha(\rho - \rho_h + \Delta\rho)\right)} \times \frac{1}{1+\exp \left(- \beta_1 \big(\frac{\rho_n}{\rho} - c_1 \big)\right)}\nonumber\\
&& k_a = \nonumber\\
&& \frac{\bar{k}_1}{1+\exp\left(-\alpha(\rho - \rho_h - \Delta\rho)\right)} + \frac{\bar{k}_2}{1+\exp\left(\beta_2 \big(\frac{\rho_n}{\rho} - c_2 \big) \right)}.\nonumber\\
\end{eqnarray}
Here, $\bar{k}_1$ tunes the amplitude of cell division and apoptosis
in the system as functions of cell density, as $\bar{k}_2$ tunes how
strongly cells die when deprived of nutrients. The parameter $\alpha$
tunes how sharply cell start to die or proliferate as the
homeostatic density is passed. It is the same in both functions
$k_d$ and $k_a$, such that $k_d-k_a=0$ at $\rho=\rho_h$
for large concentrations of nutrients. $\Delta\rho$ sets the amount
of cell turnover at homeostatic density. Finally, $\beta_1$ and
$\beta_2$ tune how sharply the cell division and apoptosis rates
change as the critical concentrations of nutrients per cell $c_1$
and $c_2$ are passed. We illustrate the dependence of $k_d$ and
$k_a$ on $\rho$ and $\rho_n$ in Fig.~\ref{fig_parametric2}, with
parameters given in Tables \ref{parametersSpherical} and
\ref{parametersNutrients}.
 Since only the difference $k_d-k_a$
enters the growth dynamics (Eq.~\textbf{1}-\textbf{4}, main text),
we illustrate the dependence of this combination on cell density and
nutrient concentration in Figs. \ref{fig_parametric} and
\ref{fig_parametric1}.

To describe how nutrients are distributed in the system, we suppose
that they diffuse freely throughout the system with a given
diffusion constant $D_n$, while being consumed by living cells for
their metabolism and their growth. Metabolism uptake happens at a
given rate $\mu\rho_n$ that is proportional to the available
concentration of nutrients, and growth dependence is described via
an extra consumption term proportional to the number of cell
division $k_d\rho$ with a coupling constant $\lambda$. For
simplicity, we suppose no effect of cell apoptosis on nutrient
uptake. We finally suppose that nutrient diffusion is very fast
compared to tissue growth, such that only the steady-state diffusion
equation needs to be considered:
\begin{equation}\label{NutrientEquation}
D_n \frac{1}{r^2} \frac{\partial}{\partial r} \left(r^2 \frac{\partial \rho_n}{\partial r} \right) - \lambda k_d \rho - \mu  \rho_n \rho = 0.
\end{equation}
Boundary conditions are as follows: the nutrient concentration is
homogeneous and constant in the healthy compartment and the
flow of nutrients vanishes at the center of the tumor compartment.

\begin{figure}[h]
\scalebox{0.19}{
\includegraphics{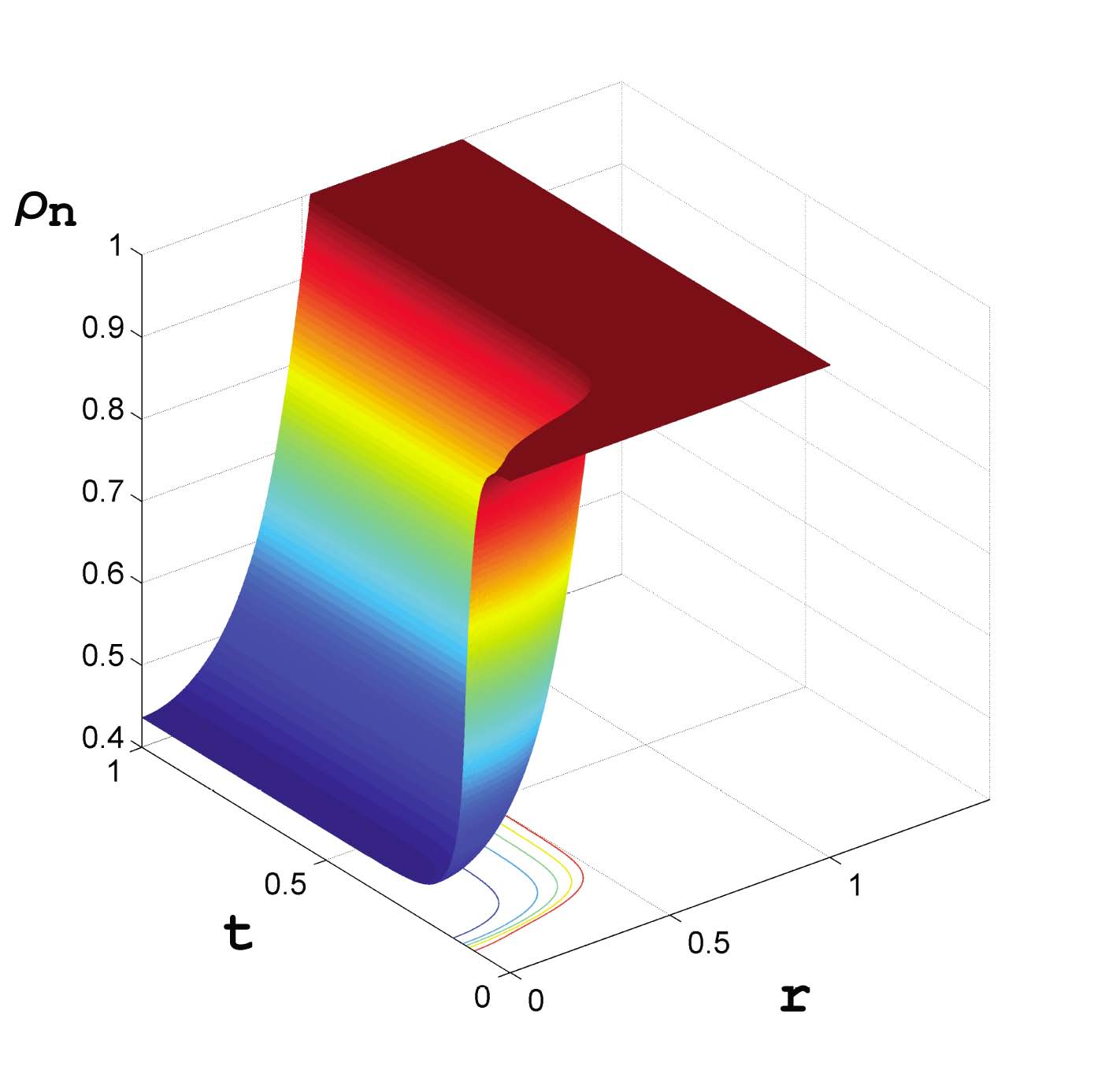}
}
\caption {\label{fig_nutconc} Typical nutrient profile as a function
of time in a growing tumor that arrests in a dormant state. The flux
at the center vanishes and the outside compartment has a constant
concentration. Parameters correspond to those given in Tables
\ref{parametersSpherical} and \ref{parametersNutrients}.}
\end{figure}

To compute the growth dynamics of a spherical tumor coupled to
nutrient diffusion through its surface, we use the same method as in
Section I, while solving the steady-state diffusion equation
(\ref{NutrientEquation}) at every timestep. The result with the
parameters given in Table \ref{parametersNutrients} is shown in
Fig.~\textbf{4} of the main text. The characteristic nutrient profile in
a tumor that is nutrient-limited in its growth is given in
Fig.~\ref{fig_nutconc}.

\section{Stochastic Dynamics}

We solve the master equation (Eq.~\textbf{5}, main text) together with the rates given by
Eq.~\textbf{7} (main text), both analytically and numerically.
Imposing an upper sink at $n_{\rm{max}}$ in addition to the lower
sink at $n_{\rm{min}}=0$, all clusters of cells end up in one of the
two sinks when time goes to infinity. The analytic solution---given by
Eq.~\textbf{8} (main text)---gives the splitting probability, namely
the probability for a cluster originating from a single cell to
reach the upper sink $n_{\rm{max}}$ when time goes to infinity.
Numerically, we use a Monte Carlo simulation of the master equation \textbf{5} (main text)
based on the Gillespie algorithm (Gillespie, 1976), which gives information on the
temporal evolution of the process leading to the analytic result
Eq.~\textbf{8} (main text) for long evolution times.

The Monte Carlo simulation is implemented according to the following
standard procedure: at each step of the growth process
(corresponding to a new division or apoptosis event), the algorithm
first sets the time delay between this event and the previous one,
and then chooses whether apoptosis or division takes place. These
two choices are made by generating two random numbers $x_1$ and
$x_2$ in the interval $[0,1]$ using a Mersenne Twister algorithm
(Matsumoto and Nishimura, 1998). $x_1$ determines the time delay $\delta t$
between two events as:
\begin{equation}
\delta t = - \frac{\log(x_1)}{r_n^+ + r_n^-},
\end{equation}
where $r_n^+$ and $r_n^-$ are the growth and death rates given
by Eq.~\textbf{7} (main text). The stochastic variable $x_2$
determines whether growth or recession takes place: growth is chosen if $[(r_n^+ + r_n^-)x_2]
\leq r_n^+$, recession otherwise.

Parameters are chosen as follows: the upper sink is at
$n_{\rm{max}}=10^6$ cells. The parameters that enter the expression
for the rates (Eq.~\textbf{7}, main text) are such that, at very
large radii, the tumor divides at a rate of one division per day on
average, while having a vanishing probability to shrink. This
yields:
\begin{equation}
\kappa_{d/a}\,\chi_{T} = \frac{k^{+/-}_{n \rightarrow \infty}-k_0}{p_{T,h}-p_{H,h}},
\end{equation}
with
\begin{eqnarray}
k^+_{n \rightarrow \infty}&=&\lim_{n \rightarrow \infty}\frac{r^+_n}{n}=1\nonumber\\
k^-_{n \rightarrow \infty}&=&\lim_{n \rightarrow \infty}\frac{r^-_n}{n}=0
\end{eqnarray}
in invert units of days. Here, $k_0$ is an adjustable parameter in
the interval $[k^-_{n \rightarrow \infty},k^+_{n \rightarrow
\infty}]$ that tunes the amount of stochasticity in the system.
Finally, we impose a critical radius $r_c$ corresponding to a
critical number of cells $n_c$ at density $\rho_{T,h}$, and choose
the surface tension accordingly as:
\begin{equation}
\gamma = \frac{p_{T,h}-p_{H,h}}{2}\left(\frac{3  n_c}{4  \pi  \rho_{T,h}}\right)^{1/3}.
\end{equation}
The parameters used in Fig.~\textbf{6} of the main text are given in Table \ref{parametersGillespie}.

\begin{widetext}

\section*{Tables}

\begin{table}[h]
\begin{tabular}{|l|l|l|l|}
\hline
Parameter & Value & Description\\
\hline
$x_0$ & $0.01$ & initial interface location\\
$\rho_0$ & $1$ & initial density\\
$v_0$ & $0$ & homogeneous initial velocity\\
\hline
$t_{\rm{evol.}}$ & $650$ & total time of evolution in Fig.~\textbf{2} (main text)\\
 & $600$ & total time of evolution in Figs. ~\textbf{3} and \textbf{5} (main text)\\
 & $3000$ & total time of evolution in Fig.~\textbf{4} (main text) and Fig.~5 here\\
\hline
$\kappa$ & $10$ & division constant (both tissues)\\
$\chi$ & $0.2$ & compressibility (both tissues)\\
$\eta$ & $50$ & viscosity (both tissues)\\
\hline
\end{tabular}
\caption{ \label{parametersSpherical} Parameters used to compute the
growth dynamics of spherical tumors in Figs. \textbf{2}, \textbf{3}, \textbf{4} and
\textbf{5} of the main text and Fig.~5 here. Parameters are
scaled in units of the total compartment size $R$, the homeostatic
pressure difference $\Delta p$ between the two tissues and the
homeostatic densities $\rho_h$ which we assume to be identical for the two
tissues. Total times are scaled to one in the figures.}
\end{table}

\begin{table}[h]
\begin{tabular}{|l|l|l|l|}
\hline
Parameter & Value & Description\\
\hline
$R$ & $100$ $\mu$m & total compartment radius (Weinberg, 2007)\\
$x_0$ & $10$ $\mu$m & initial interface location \\
$\rho_h$ & $0.001$ $\mu$m$^3$ & homeostatic density (both tissues) (Weinberg, 2007)\\
$\Delta p$ & $1000$ Pa & difference in homeostatic pressures (both tissues) (Helminger {\it et al.}, 1997)\\
$\chi $ & $10^{-7}$ Pa$^{-1}$$\cdot$$\mu$m$^{-3}$ & compressibility (both tissues) (Tschumperlin  {\it et al.}, 2004)\\
$\kappa \cdot \chi \cdot \Delta p$ & $1$ d$^{-1}$ & maximum division rate (both tissues) (Weinberg, 2007)\\
$\eta$ & $10^4$ Pa$\cdot$s & tissue viscosity (Forgacs {\it et al.}, 1998)\\
\hline
\end{tabular}
\caption{
\label{parametersRealistic}
List of parameters used to compute the growth dynamics of spherical tumors in Fig.~1.
}
\end{table}

\begin{table}[h]
\begin{tabular}{|l|l|l|l|}
\hline
Parameter & Value & Description\\
\hline
$c_1$ & $0$ & nutrients per cell for induction of apoptosis\\
$c_2$ & $0.6$ & nutrients per cell for arrest of proliferation\\
$\bar{k}_1$ & $10$ & maximum cell division and apoptosis at high nutrient concentration\\
$\bar{k}_2$ & $500$ & apoptosis rate coefficient at starvation\\
$\alpha$ & $10$ & response coefficient of cell division and apoptosis to cell density\\
$\beta_1$ & $50$ & response coefficient of cell division to nutrient concentration per cell\\
$\beta_2$ & $500$ & response coefficient of cell apoptosis to nutrient concentration per cell\\
$\Delta \rho$ & $0.1$ & shift in cell division and apoptosis tuning the amount of\\
 &  & cell turnover at homeostatic density\\
\hline
$D_n$ & $1$ & nutrient diffusion constant\\
$\lambda$ & $6$ & nutrient consumption for proliferation\\
$\mu$ & $1000$ & nutrient consumption for metabolism\\
\hline
\end{tabular}
\caption{ \label{parametersNutrients} Parameters used to generate
the plots of Fig.~\textbf{4} of the main text and Figs. 2-5 here in the
coupling of the growth dynamics to nutrients as they enter in
Eq.~(\ref{rates}).}
\end{table}

\begin{table}[h]
\begin{tabular}{|l|l|l|l|}
\hline
Parameter & Value & Description\\
\hline
$k^+_{n \rightarrow \infty}$ & $1$ d$^{-1}$ & $n$ infinity value of $r^+_n/n$\\
$k^-_{n \rightarrow \infty}$ & $0$ d$^{-1}$ & $n$ infinity value of $r^-_n/n$\\
$k_0$ & $0.9$ d$^{-1}$ & cell turnover at homeostatic density\\
\hline
\end{tabular}
\caption{
\label{parametersGillespie}
Parameters used in Fig.~\textbf{6}  of the main text.}
\end{table}

\end{widetext}

\end{document}